\documentclass[aps,twocolumn,groupedaddress,showpacs]{revtex4}
\usepackage{amsmath}
\usepackage{graphicx}
\begin{document}
\bibliographystyle{apsrev}

\title{Conductance Fluctuations and Partially Broken Spin Symmetries in Quantum Dots}
\author{D.~M.~Zumb\"uhl}
\affiliation{Department of Physics, Harvard University,
Cambridge, Massachusetts 02138}
\author{J.~B.~Miller}
\affiliation{Department of Physics, Harvard University,
Cambridge, Massachusetts 02138}
\author{D.~Goldhaber-Gordon}
\affiliation{Department of Physics, Stanford University,
Stanford, California 94305}
\author{C.~M.~Marcus}
\affiliation{Department of Physics, Harvard University,
Cambridge, Massachusetts 02138}
\author{J.~S.~Harris, Jr.}
\affiliation{Departement of Electrical Engineering,
Stanford University, Stanford, California 94305}
\author{K.~Campman}
\affiliation{Materials Department, University of
California, Santa Barbara, California 93106}
\author{A.~C.~Gossard}
\affiliation{Materials Department, University of
California, Santa Barbara, California 93106}


\begin{abstract}
Conductance fluctuations in GaAs quantum dots with
spin-orbit and Zeeman coupling are investigated
experimentally and compared to a random matrix theory
formulation that defines a number of regimes of spin
symmetry depending on experimental parameters. Accounting
for orbital coupling of the in-plane magnetic field, which
can break time-reversal symmetry, yields excellent overall
agreement between experiment and theory.
\end{abstract}

\maketitle

The combination of confinement, spin-orbit (SO) coupling,
and Zeeman splitting in semiconductor quantum dots gives
rise to rich physics, including experimental access to
interesting partially-broken spin symmetries
\cite{Aleiner01} and a suppression of SO effects due to
confinement \cite{Khaetskii00, Halperin01, Aleiner01,
Zumbuhl02, Folk01} that provides long spin lifetimes in
small quantum dots \cite{Khaetskii00, Fujisawa02a}. Further
consequences of these combined effects are that the
confinement-induced suppression of SO effects is lifted by
adding a Zeeman field \cite{Halperin01, Aleiner01, Folk01}
or by allowing spatial dependence of the SO coupling
\cite{Brouwer03}. Because of the finite thickness of a
two-dimensional electron gas, an in-plane magnetic field
$B_\parallel$ will have an orbital coupling that affects
the electron dispersion and can break time-reversal
symmetry (TRS) \cite{Falko02, Meyer02, Zumbuhl02,
Zumbuhl04, SOBparBulk, Meijer04}, adding additional
complexity to this system.

This Letter presents an experimental study of mesoscopic
conductance fluctuations in quantum dots that possess both
significant SO and Zeeman coupling. We find that the
$B_\parallel$ dependence of the variance of conductance
fluctuations, var $g$, with TRS explicitly broken by a
perpendicular field ($B_\bot \ne 0$, i.\,e., $B_\bot \gg
(h/e)/A$, where A is the dot area) depends critically on
the strength of SO coupling and dot size. This dependence
can be understood in terms of spin symmetries partially
broken by $B_\parallel$ and is in quantitative agreement
with an appropriate random matrix theory (RMT) formulation
\cite{Aleiner01}. We also find that var $g(B_\bot,
B_\parallel)$ becomes independent of $B_\bot$ at large
$B_\parallel$ due to $B_\parallel$ breaking TRS, consistent
with previous results \cite{Zumbuhl02, Zumbuhl04}. Taking
into account orbital coupling \cite{Meyer02, Falko02},
agreement between theory and experiment is excellent for
both broken and unbroken TRS and various regimes of spin
symmetry.

In quantum dots, effects of Rashba and linear Dresselhaus
SO coupling are suppressed due to confinement in the
absence of Zeeman coupling
\cite{Khaetskii00,Halperin01,Aleiner01}. For large Zeeman
splitting or weak confinement, this suppression is lifted
and new symmetry classes with partially broken spin
symmetry appear.  A random matrix theory (RMT) analysis of
this system was developed by Aleiner {\it et al}.\
\cite{Aleiner01} and extended to include inhomogeneous SO
coupling and interpolation between ensembles
\cite{Aleiner01}.  The RMT formulation identifies three
symmetry parameters that govern the amplitude (variance) of
conductance fluctuations, $\mathrm{var}\, g\propto s/(\beta
\Sigma)$. Here, $\beta = \{1,2,4\}$ is the usual Dyson
parameter reflecting TRS, $s = \{1,2\}$ accounts for
Kramers degeneracy, and $\Sigma= \{1,2\} $ characterizes
mixing between Kramers pairs. With these parameters, spin
symmetry may be either unbroken ($s=2, \Sigma=1$),
partially broken ($s=1, \Sigma=1$) or fully broken ($s=1,
\Sigma=2$), causing a reduction of the variance by a factor
of two each time spin symmetry is incrementally broken.
Temperature and decoherence also reduce $\mathrm{var}\,
(g)$, but ratios such as $\mathrm{var}\,
g(B_\parallel)/\mathrm{var}\, g(B_\parallel=0)$ are
affected only weakly.

Conductance fluctuations are known to be reduced by SO and
Zeeman coupling in bulk (disordered) mesoscopic samples,
and theories \cite{Meir89, VarSOBpar, Lyanda-Geller94} are
in good agreement with experiments \cite{BulkExp}.
Recently, the combined effects of SO and Zeeman coupling on
magneto resistance in bulk samples were investigated
\cite{SOBparBulk}, reporting spin-induced breaking of TRS
\cite{Meijer04}. Experimental observation of partially
broken spin symmetry, which has been theoretically
predicted \cite{Lyanda-Geller94, Aleiner01}, has to our
knowledge not been previously reported. Results of
Ref.~\cite{Aleiner01} were used to explain existing data on
var~$g(B_\bot\neq 0,B_\parallel)$ \cite{Folk01} as well as
subsequent experiments on average conductance (weak
localization and antilocalization) in quantum dots
\cite{Zumbuhl02, Hackens02}. There have been no comparable
studies of conductance fluctuations to investigate the
various symmetry classes to our knowledge.

\begin{figure}[&t]
\includegraphics[width=2.9in]{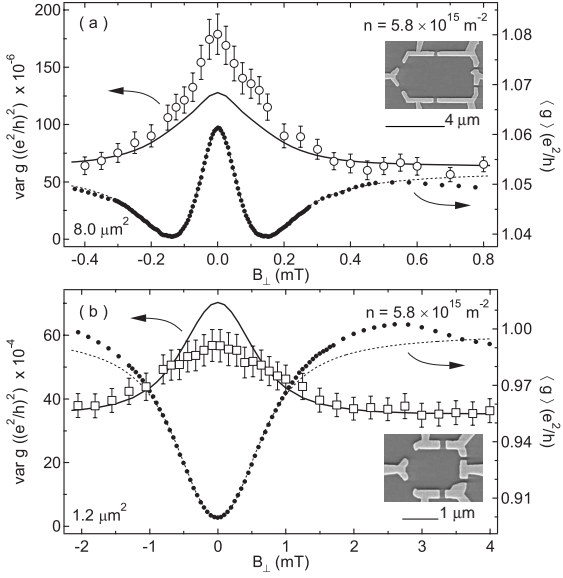}
\caption{\footnotesize{\label{VarF1} Conductance average
$\langle g(B_\bot) \rangle$ (solid dots) and variance
$\mathrm{var}\, g(B_\bot)$ (open circles and squares) as a
function of perpendicular magnetic field $B_\bot$ with
$B_\parallel=0$, for large and small dots on the high
density material, at $T=300 $ mK. (a) Antilocalization in
$\langle g(B_\bot) \rangle$ for the $8\, \mathrm{\mu m^2}$
dot. (b) Weak localization in $\langle g(B_\bot) \rangle$
for the $1.2\, \mathrm{\mu m^2}$ dot, demonstrating
confinement suppression of SO effects. Both dots show an
enhancement of var $g$ at $B_\bot=0$. Fits of RMT
\cite{Aleiner01} to $\langle g (B_\bot) \rangle$ (dashed
curves) and $\mathrm{var}\, g(B_\bot)$ (solid curves) using
fit parameters determined from fits to $\langle g \rangle$
plus an overall scale factor for var $g$ (see text). Insets
show device micrographs.}}
\end{figure}

Four gate-defined quantum dots of various sizes on two
heterostructure wafers were measured (see Table 1 and
Figs.\ 1 and 2 insets). The lower density wafer showed weak
localization at $B_{\parallel}=0$ \cite{Zumbuhl04}, while
the higher density material has sufficient SO coupling to
exhibit antilocalization at $B_{\parallel}$
\cite{Zumbuhl02}. Further details of these wafers are given
in \cite{Zumbuhl02, Zumbuhl04}. Measurements were made in a
$ ^3$He cryostat at $0.3\, \mathrm{K}$ using current bias
of $1 \, \mathrm{nA}$ at $338\,\mathrm{Hz}$. In order to
apply tesla-scale $B_\parallel$ while maintaining sub-gauss
control of $B_\perp$, we mount the sample with the 2DEG
aligned to the axis of the primary solenoid (accurate to
$\sim 1^\circ$) and use an independent split-coil magnet
attached to the cryostat to provide $B_\perp$
\cite{Folk01}. The Hall voltage measured in a co-mounted
Hall bar sample as well as the symmetry of transport
through the dot itself (visible for $B_\parallel \lesssim 2
T$) was used to locate $B_\bot=0$ as it changed with
$B_\parallel$.

\begin{figure}[t]
\includegraphics[width=2.9in]{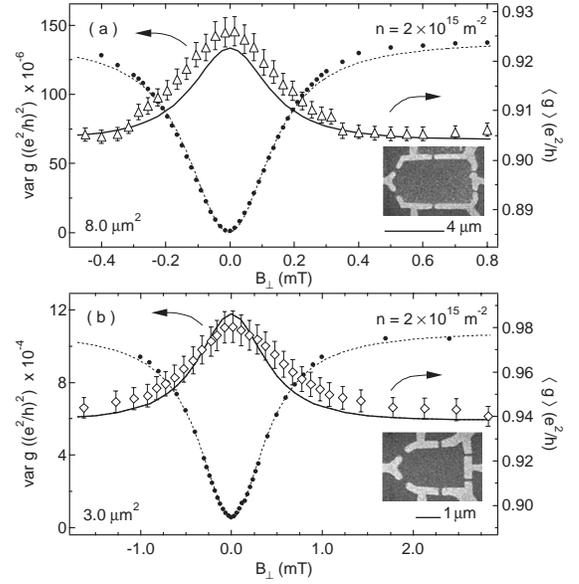}
\caption{\footnotesize{\label{VarF2} Conductance average
$\langle g(B_\bot) \rangle$ (solid dots) and variance
$\mathrm{var}\, g(B_\bot)$ (open triangles and diamonds) as
a function of perpendicular field $B_\bot$ with
$B_\parallel=0$, for (a) large and (b) small dot, both
fabricated on low density material. Both devices display
weak localization in $\langle g(B_\bot) \rangle$. Fits to
RMT is shown as dashed and solid curves, as described in
the caption of Fig.~1. Insets show device micrographs;
geometry of large device is identical to large dot in high
density material.}}
\end{figure}

Statistics of conductance fluctuations were gathered using
two shape-distorting gates \cite{Chan95} while the point
contacts were actively held at one fully transmitting mode
each. At each value of $B_\bot$ and $B_\parallel$, mean and
variance were estimated based on $\sim 400$ ($\sim 200$)
statistically independent samples for the low density (high
density) dots. For var $g(B_\parallel, B_\bot\neq 0)$ data
with TRS explicitly broken, $B_\bot$ was used to gather
additional samples to reduced the statistical error.

\begin{figure}[t]
\includegraphics[width=2.9in]{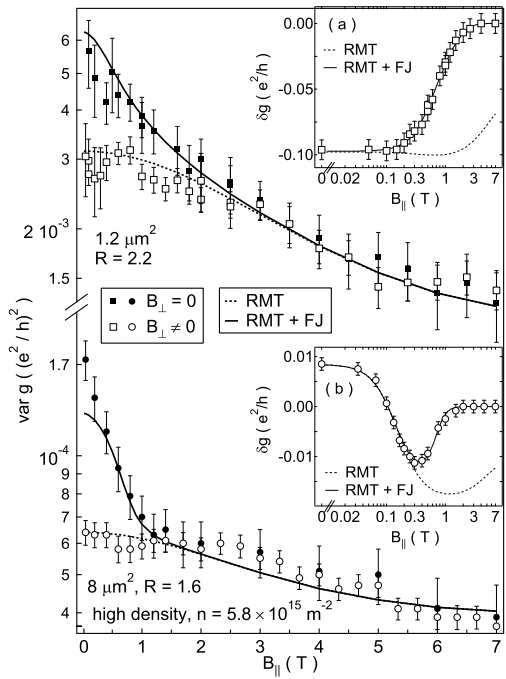}
\caption{\footnotesize{\label{VarF3} Variance of
conductance fluctuations, var $g$, in high density dots, as
a function of in-plane field $B_\parallel$ with $B_\bot\neq
0$ sufficient to break TRS (open symbols) and $B_\bot=0$
(solid symbols). The big dot shows less reduction in var
$g(B_\bot\neq 0)$ with $B_\parallel$ than the small dot,
consistent with RMT, see text. Insets show quantum
correction to average conductance, $\delta
g(B_\parallel)=\langle g(B_\bot=0,B_\parallel)\rangle
-\langle g(B_\bot\neq 0, B_\parallel)\rangle$. In both main
figure and insets, dashed curves are fits to RMT, solid
curves (labeled RMT+FJ) are fits to RMT including orbital
coupling of $B_\parallel$, see text.}}
\end{figure}

\begin{table}[&b]
\label{table1}
\begin{tabular}{c|c|c|c|c|c|c|c|c|c|c}
\rule[-2mm]{0mm}{3mm} $n$ & $A$ & $\tau_\varphi$ & $\lambda_{so}$ & $\nu_{so}$ &
$\kappa_\bot$ & $f_{var}$ & $\xi$ & a & b\\
$\mathrm{m^{-2}}$ & $\mathrm{\mu m^2}$& ns & $\mathrm{\mu m}$ & & & & &
$\mathrm{ns^{-1}T^{-2}}$ & $\mathrm{ns^{-1}T^{-6}}$\\
\hline 2.0 &  3.0 & 0.18 & 8.5 & 1.0 & 0.15 & 1.0 & 2.8  & 0.5$\pm$0.1 & 0.028
\\
2.0 &  8.0 & 0.21 & 8.5 & 1.0 & 0.25 & 0.6 & 3.0  & 0.37$\pm$0.07 & 0.028  \\
5.8 &  1.2 & 0.10 & 3.2 & 1.4 & 0.33 & 1.9 & 1.0  & 6.6$\pm$1 & 0.14  \\
5.8 &  8.0 & 0.39 & 4.4 & 1.4 & 0.23 & 0.7 & 0.45 & 1.4$\pm$0.4 & 0.14  \\
\end{tabular}
\caption{\footnotesize{Carrier density $n$, dot area $A=L^2$,
coherence time $\tau_\varphi$, spin-orbit parameters
$\lambda_{so}$ and $\nu_{so}$}, RMT parameters $\kappa_\bot$,
$f_{var}$ and $\xi$ and FJ parameters $a$ and $b$, see text.}
\end{table}

Fitting the RMT results to $\langle g(B_\bot) \rangle$
yields values for the average SO length
$\lambda_{so}=\sqrt{|\lambda_1 \lambda_2|}$, where
$\lambda_{1,2}$ are the SO lengths along the main crystal
axes, as well as the phase coherence time $\tau_\varphi$
and a geometrical parameter $\kappa_\bot$. The SO
inhomogeneity $\nu_{so}=\sqrt{|\lambda_1/\lambda_2|}$ can
be extracted from $\langle g(B_\parallel) \rangle$ in the
presence of antilocalization (AL), and is taken as
$\nu_{so}=1.4(1.0)$ for the high (low) density devices
\cite{Zumbuhl02}. An additional order-one geometrical
parameter $\kappa'$, relevant for the strong SO regime and
not readily extracted from transport measurements, is set
to $\kappa' = 1$ for all devices. Further details of fits
to the average conductance are given in
Ref.~\cite{Zumbuhl02}. We note that in absence of AL
$\lambda_{so}$ can only be bounded from below. Values for
$\tau_\varphi$ are similar for all devices and consistent
with previous measurements \cite{Huibers99}.

As seen in Figs.~\ref{VarF1} and \ref{VarF2}, all devices
except the large high density dot show weak localization
(WL), indicating these dots are in the regime $\lambda_{so}
\gg L$, where SO effects are strongly suppressed by
confinement. The observation of WL down to $\sim 40\,
\mathrm{mK}$ in previous measurements on identical devices
\cite{Huibers99} bounds the SO rate to $\lambda_{so}\gtrsim
9 \, \mathrm{\mu m}$. We note that SO coupling of this
order will noticeably reduce the low-temperature WL feature
and therefore contribute to the saturation of WL observed
in Ref.~\cite{Huibers99}.

\begin{figure}[t]
\includegraphics[width=2.9in]{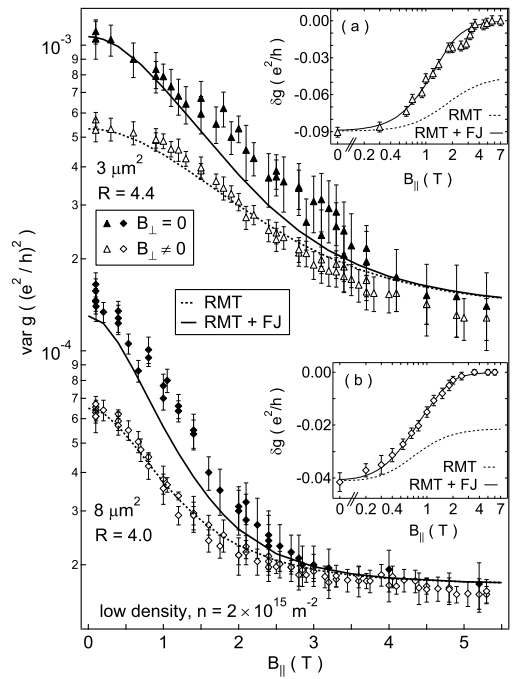}
\caption{\footnotesize{\label{VarF4} As in Fig.~3, but for
low density dots. In this case, the larger dot has a {\em
larger} reduction factor in parallel field, consistent with
RMT (curves). Orbital coupling of $B_\parallel$ breaks TRS,
making var $g$ independent of $B_\bot$ and quenching the
quantum correction to average conductance, $\delta g
\rightarrow  0$, (insets) on the same (dot-size-dependent)
scale of $B_\parallel$}}
\end{figure}

For both the high and low density material (respectively
with AL and WL), var $g$ at $B_\parallel=0$ is reduced upon
application of a TRS-breaking perpendicular field, as seen
in Figs.~\ref{VarF1} and \ref{VarF2}. This effect has been
investigated previously for the weak SO (WL) case
\cite{Chan95, Huibers98b}, and has been observed but not
analyzed for the strong SO (AL) case \cite{Zumbuhl02}. The
solid theory curves in Figs.~\ref{VarF1}, \ref{VarF2}
include thermal smearing and decoherence effects and use
parameters obtained from fits of RMT to $\langle
g(B_\bot)\rangle$, plus one additional parameter,
$f_{var}$, (Table 1) to normalize the var $g(B_\bot\neq 0)$
the RMT value. This factor compensates the assumption of
multimode leads,  $N \gg 1$, in the RMT \cite{Aleiner01},
whereas the experiment has $N=2$. RMT for var~$g$ in the
case $N=2$ case has been given, but does not include SO or
Zeeman terms \cite{Alves02}.

We next investigate the effect of an in-plane magnetic field on var $g$,
focusing first on the case where TRS is broken by a small perpendicular
field, $B_\bot\neq0$. As seen in Figs.~\ref{VarF3} and \ref{VarF4}, var
$g(B_{\bot} \ne 0, B_{\parallel})$ decreases with increasing
$B_{\parallel}$, and saturates at large $B_\parallel$, giving reduction
factors $R=\mathrm{var}\, g(B_\bot\neq 0,B_\parallel=0)/\mathrm{var}\,
g(B_\bot\neq 0, B_\parallel \gg 0)$ between $R\sim1.6$ for the large
high-density dots (which show AL at $B_{\parallel}=0$) and $R\sim4$ for
large low-density dots (which show WL at $B_\parallel=0$). This range of
values for $R$ can be readily interpreted within RMT: For the large high
density dots (relatively strong SO coupling, not suppressed by
confinement), Kramers degeneracy is broken whenever TRS is broken, and
there is weak spin mixing ($\beta = 2, s=1, 1 < \Sigma < 2$) at
$B_{\parallel} = 0$. The effect of $B_{\parallel}$ is to fully mix the
Kramers pair ($ \beta = 2, s=1, \Sigma = 1$), thus the reduction $1\leq
R\leq 2$. On the other hand, dots showing WL at $B_\parallel=0$ retain
spin degeneracy ($\beta = 2, s = 2, \Sigma = 2$) at $B_\parallel=0$, which
is then lifted by Zeeman coupling ($\beta = 2, s = 1, \Sigma = 2$) and at
larger $B_\parallel$ mixed ($\beta = 2, s = 1, \Sigma = 1$) due to SO
coupling revived by $B_\parallel$, giving $R\sim4$.

Spin mixing induced by $B_\parallel$, marking the  $\Sigma = 1$
to $2$ crossover, occurs when a field-dependent energy scale
$\epsilon_\bot^Z$ exceeds the level broadening $\tilde\gamma =
\hbar (\tau_{esc}^{-1} + \tau_\varphi^{-1})^{-1}$
($\tau_{esc}^{-1}=N\Delta/h$ is the escape rate from the dot).
This new energy scale depends on both Zeeman and SO coupling,
$\epsilon_\bot^Z=\xi^2 \epsilon_Z^2/(2 E_T) (A/\lambda_{so}^2)$
\cite{Aleiner01}, where $\epsilon_Z=g\mu_B B$ is the Zeeman
splitting, $E_T$ is the Thouless energy (for ballistic dots
$E_T\sim \hbar v_F /\sqrt{A}$, where $v_F$ is the Fermi
velocity), and $\xi$ is a parameter of order one that depends on
dot geometry as well as the direction of $B_\parallel$ \cite
{Halperin01, Aleiner01}. Note that $\epsilon_\bot^Z/\tilde\gamma
\propto \epsilon_Z^2 A^{5/2}$ (when $\tau_{esc}\ll
\tau_\varphi$) so that the $\Sigma = 1$ to $2$ crossover field
will depend on dot size. For the smallest dot, the crossover is
inaccessible, and $\Sigma = 1$ for all measured fields and
$R\sim2$ due to breaking of Kramers degeneracy only ($s = 2$ to
$1$). In the low density dots, the $\Sigma = 1$ to $2$ crossover
is accessible, occurring around $B_\parallel \sim 1(3)\,
\mathrm{T}$ for the larger (smaller) dot. The large high density
dot has $1<\Sigma<2$ already at $B_\parallel=0$ due to SO
coupling, and the crossover to $\Sigma=2$ occurs around
$B_\parallel \sim 3\, \mathrm{T}$. Because of the undetermined
coefficient $\xi$, the SO length $\lambda_{so}$ cannot be
independently extracted from $\mathrm{var}\, g(B_\parallel)$.
Taking $\xi$ as a single fit parameter, the dashed curves in
Figs.~\ref{VarF3} and \ref{VarF4} give the RMT results, which
are in good agreement with experiment for all devices.

Finally, we investigate orbital effects of $B_\parallel$ on
var~$g$, measured when TRS is not explicitly broken by a
perpendicular field ($B_\bot=0)$. Figures~\ref{VarF3} and
\ref{VarF4} show that as $B_\parallel$ is increased,
$\mathrm{var}\, g(B_\bot=0,B_\parallel)$ decreases sharply,
approaching $\mathrm{var}\, g(B_\bot\neq 0,B_\parallel)$.
At large $B_\parallel$, var~$g$ becomes independent of
$B_\bot$ whereas RMT gives
var~$g_{RMT}(B_\bot=0)/$var~$g_{RMT}(B_\bot \neq 0) = 2$
for all $B_\parallel$. On a similar scale of $B_\parallel$,
WL corrections $\delta g(B_\parallel)=\langle
g(B_\bot=0,B_\parallel)\rangle -\langle g(B_\bot\neq 0,
B_\parallel)\rangle$ are also vanishing in all devices
whereas RMT predicts a finite $\delta g$. As discussed
previously \cite{Folk01, Zumbuhl02, Zumbuhl04}, these
effects result from the breaking of TRS by $B_\parallel$
\cite{Meyer02, Falko02}.

Following Ref.~\cite{Falko02} (FJ), we account for the
suppression of $\delta g(B_\parallel)$ and var $g(B_\parallel)$
by introducing a field-dependent factor
$f_{FJ}(B_\parallel)=(1+\tau_{B\parallel}^{-1}/\tau_{esc}^{-1})^{-1}$,
where $\tau_{B\parallel}^{-1}\sim
aB_\parallel^2+bB_\parallel^6$. The $B_\parallel^2$ term
reflects interface roughness and dopant inhomogeneities; the
$B_\parallel^6$ term is due to the asymmetry of the well. The
RMT results are then modified as $\delta g(B_\parallel)=\delta
g_{RMT}(B_\parallel)f_{FJ}(B_\parallel)$ and $\mathrm{var}\,
g(B_\bot=0,B_\parallel)=\mathrm{var}\, g_{RMT}(B_\bot\neq0,
B_\parallel)(1+f_{FJ}(B_\parallel))$ to account for flux effects
of the parallel field \cite{FalkoComm}. The coefficient $a$ is
obtained from a fit to the experimental $\delta g(B_\parallel)$
while $b$ is estimated from device simulations \cite{FalkoComm}
(Table I). The resulting theory curves for both $\delta
g(B_\parallel)$ (solid curves, insets Figs. \ref{VarF3} and
\ref{VarF4}) and $\mathrm{var}\, g(B_\bot=0,B_\parallel)$ (solid
curves, main panels) are in good agreement with experiment. We
emphasize that the theoretical variance curves $\mathrm{var}\,
g(B_\bot=0,B_\parallel)$ are not fit. Estimates of $a,b$ based
on correlation functions of parallel-field conductance
fluctuations \cite{Zumbuhl04} are consistent with the values
obtained here based on $\delta g(B_\parallel)$.

In summary, mesoscopic conductance fluctuations in open quantum
dots in presence of SO coupling and in-plane fields can be
understood in terms of fundamental symmetries in the system,
including novel partially broken spin rotation symmetries as
well as time reversal symmetry, which can be broken by both
perpendicular and in-plane fields.

We thank I. Aleiner, B. Altshuler, P. Brouwer, J. Cremers,
V. Fal'ko, J. Folk, B. Halperin, T. Jungwirth and Y.
Lyanda-Geller. This work was supported in part by
DARPA-QuIST, DARPA-SpinS, ARO-MURI and NSF-NSEC. We also
acknowledge support from NDSEG (J.B.M.) and the Harvard
Society of Fellows (D.G.-G). Work at UCSB was supported by
QUEST, an NSF Science and Technology Center.


\end{document}